\documentclass[aps,prb,twocolumn,showpacs]{revtex4}
\usepackage{color}
\usepackage{graphicx}
\usepackage{amsmath}
\usepackage{amssymb}
\usepackage{amsfonts}
\usepackage[colorlinks,hyperindex]{hyperref}

\begin{document}
\title{Bright sink-type localized states in exciton-polariton condensates}
\author{Micha{\l} Kulczykowski, Nataliya Bobrovska, and Micha{\l} Matuszewski}
\affiliation{
Institute of Physics Polish Academy of Sciences, Al. Lotnik\'ow 32/46, 02-668 Warsaw, Poland
}

\begin{abstract}
The family of one-dimensional localized solutions to dissipative nonlinear equations includes a variety of objects such as sources, sinks, shocks (kinks), and pulses. 
These states are in general accompanied by nontrivial density currents, which are not necessarily related to the movement of the object itself. 
We investigate the existence and physical properties of sink-type solutions in nonresonantly pumped 
exciton-polariton condensates modeled by an open-dissipative Gross-Pitaevskii equation. 
While sinks possess density profiles similar to bright solitons,
they are qualitatively different objects as they exist in the case of repulsive interactions and represent a heteroclinic solution.
We show that sinks can be created in realistic systems with appropriately designed pumping profiles. 
We also consider the possibility of creating sinks in a two-dimensional configuration with a ring-shaped pumping profile.
\end{abstract}
\maketitle

\section{Introduction}

Strong coupling of semiconductor excitons to microcavity photons results in the appearance
of spectral resonances associated with mixed quantum quasiparticles called 
exciton-polaritons~\cite{Polaritons}.
These particles exhibit extremely light effective masses, few orders of magnitude
smaller than the mass of electron, which allows for the observation of physical phenomena related to Bose-Einstein condensation
already at room 
temperatures~\cite{Kasprzak_BEC,Grandjean_RoomTempLasing,Kena_Organic,Mahrt_RTCondensatePolymer,
Superfluidity,Deveaud_QuantumVortices,Carusotto_QuantumFluids,Yamamoto_RMP}.
At the same time, polaritons exhibit strong exciton-mediated interparticle interactions
and picosecond lifetime due to their photonic component.
They are actively studied both from the point of view of fundamental
interest~\cite{Superfluidity,Deveaud_QuantumVortices,Carusotto_QuantumFluids}
and potential applications~\cite{Polariton_applications,ElectricallyPumped}.

In recent years, great attention has been devoted to the study of nonlinear self-localized states of superfluid polaritons, such as dark and bright 
solitons~\cite{DarkResonantSolitons,BrightResonantSolitons,Elena,Xue_DarkSolitons,Flayac_DarkSolitons,
Ostrovskaya_DarkSolitons}. Solitons are nonlinear wavepackets which preserve their shape
thanks to the balance between dispersion and nonlinearity~\cite{SomeSolitonBook}. 
They have been applied to long-distance optical fiber communication~\cite{Hasegawa} as well as description of numerous physical systems.
Polariton solitons have been demonstrated both in the cases of resonant~\cite{DarkResonantSolitons,BrightResonantSolitons} and nonresonant pumping~\cite{Elena,Xue_DarkSolitons,Flayac_DarkSolitons,
Ostrovskaya_DarkSolitons}. To date, no bright states were shown to exist in the nonresonant case with homogeneous pumping.

Polariton superfluids are inherently nonequilibrium systems 
in which the balance between pumping and loss is an essential factor~\cite{Kasprzak_BEC,Yamamoto_RMP,Wouters_ExcitationSpectrum}.
In many of the previous studies, this aspect was treated as an unwanted complication of the theory. Standard models, such as the
conservative Gross-Pitaevskii equation, were frequently used to describe solitons. 
However, it is well known that self-localized solutions in dissipative systems have qualitatively different properties 
than their conservative counterparts. In the case of repulsive interactions, only one type of 
one-dimensional solution exists
in the conservative theory -- dark or bright solitons, depending on the sign of the effective mass.
In the dissipative case, a family of qualitatively different
localized states exists, including sources, sinks, shocks and pulses~\cite{Malomed_Sinks,Hohenberg_FrontsPulses,Aranson_CGLEWorld,ConteMusette,Pastur_SourcesSinks}.
In general, they exhibit nontrivial internal density currents and may undergo complicated, 
sometimes even chaotic dynamics~\cite{Explosions,Xue_DarkSolitons}.

In this paper, we demonstrate the existence and stability of bright self-localized solutions of the open-dissipative
polariton model. These solutions are classified as sinks (anti-dark solitons)~\cite{Hohenberg_FrontsPulses,Pastur_SourcesSinks}, 
the name reflecting that their structure corresponds to terminating lines
of incoming density currents, with a local increase of loss. 
While sink-type solutions possess density profiles similar to bright solitons,
they are qualitatively different objects. In contrast to bright solitons, they exist in the case of repulsive 
interactions and represent a heteroclinic solution connecting two counterpropagating plane waves.
We demonstrate the dynamics of sink formation and their stability in a realistic model with appropriately chosen pumping profile. 
We investigate systematically the properties of sinks and provide an approximate analytical formula for their shape.
In the two-dimensional case, we show that sink creation is hindered by the spontaneous proliferation of vortices, which destroy the supercurrents necessary for the existence of a symmetric sink solution.

\section{Model}

We consider a polariton condensate in the one-dimensional (1D) setting, e.g.~trapped in a microwire~\cite{Bloch_ExtendedCondensates}. 
We model the system with the generalized open-dissipative Gross-Pitaevskii equation for the condensate wavefunction $\psi(x,t)$ coupled to 
the rate equation for the polariton reservoir density, $n_R(x,t)$ 
\cite{Wouters_ExcitationSpectrum,Xue_DarkSolitons,Bobrovska_Stability}
\begin{equation}
\begin{split}
\label{GP1}
i \hbar\frac{\partial \psi}{\partial t} &=-\frac{\hbar^2 D}{2 m^*} \frac{\partial^2 \psi}{\partial x^2} 
+ g_{\rm C}^{\rm 1D} |\psi|^2 \psi + g_{\rm R}^{\rm 1D} n_{\rm R} \psi \\
&+i\frac{\hbar}{2}\left(R^{\rm 1D} n_{\rm R} - \gamma_{\rm C} \right) \psi, \\
\frac{\partial n_{\rm R}}{\partial t} &= P(x) - (\gamma_{\rm R}+ R^{\rm 1D} |\psi|^2) n_{\rm R} 
\end{split}
\end{equation}
 where $P(x)$ is the exciton creation rate determined by the pumping profile, $m^*$ 
is the effective mass of lower polaritons, 
$\gamma_{\rm C}$ and $\gamma_{\rm R}$ are the polariton and exciton loss rates, 
and $(R^{\rm 1D},g_i^{\rm 1D})=(R^{\rm 2D},g_i^{\rm 2D})/\sqrt{2\pi d^2}$ are
the rates of stimulated scattering into the condensate and the interaction coefficients, 
rescaled in the one-dimensional case. 
Here, we assumed a Gaussian transverse profile of $|\psi|^2$ and $n_{\rm R}$ of width $d$.
In the case of a one-dimensional microwire~\cite{Bloch_ExtendedCondensates}, 
the profile width $d$ is of the order of the microwire thickness.
We also introduced $D=1-iA$ with $A$ being a small constant accounting for the 
energy relaxation in the condensate~\cite{Bobrovska_Stability,Sieberer_DynamicalCritical,Bloch_ExtendedCondensates,Bloch_GapStates}.

To obtain a system of dimensionless evolution equations, it is possible to rescale time, space, wavefunction amplitude and material coefficients 
according to $t=\tau \widetilde{t}$, $x=\xi\widetilde{x}$, $\psi=(\xi\beta)^{-1/2}\widetilde{\psi}$, $n_R=(\xi\beta)^{-1}\widetilde{n}_R$,  $R^{\rm 1D}=(\xi\beta/\tau)\widetilde{R}$, $(g^{\rm 1D}_C,g_R^{\rm 1D})=(\hbar\xi\beta/\tau)(\widetilde{g}_C,\widetilde{g}_R)$, $(\gamma_C,\gamma_R)=\tau^{-1}(\widetilde{\gamma}_C,\widetilde{\gamma}_R)$, $P(x)=(1/\xi\beta\tau)\widetilde{P}(x)$, where $\xi=\sqrt{\hbar \tau /2m^*}$, while $\tau$ and $\beta$ are arbitrary scaling parameters. We rewrite the above equation in the dimensionless form (we omit the tildes for convenience)
\begin{align}
i\frac{\partial \psi}{\partial t} &= \left[-D\frac{\partial^2}{\partial x^2} +g_C|\psi|^2 +g_Rn_R + \frac{i}{2}\left(Rn_R-\gamma_C\right)\right]\psi,\nonumber \\
\frac{\partial n_R}{\partial t} &= P(x)-\left(\gamma_R+R|\psi|^2\right)n_R.
\label{eq:PolaritonGPEWithA}
\end{align}
In the above transformations the norms of both fields $N_\psi=\int|\psi|^2 dx$ and $N_R=\int n_R dx$ are multiplied by the factor of $\beta$.

\section{Sink-type solutions}

\subsection{Description of sinks solutions}

\begin{figure}
	\begin{center}
	\includegraphics[width=1.\columnwidth]{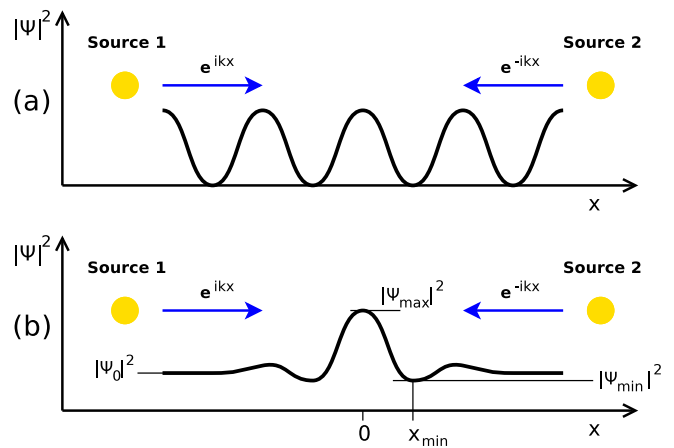}
	\caption{Density patterns created with counterpropagating waves. 
(a) Interference pattern  in the linear regime. (b) Stationary sink-type  
localized pattern in a model with nonlinear dissipation.}
	\label{fig:drawing}
	\end{center}
\end{figure}

The structure of sink solutions can be understood most easily as a result of interaction of two counterpropagating nonlinear waves.
In the linear case, two waves emitted by distant sources give rise to a standard interference pattern, as shown in Fig.~\ref{fig:drawing}(a). 
In the case of a dissipative model with nonlinear gain or loss coefficients, the interference pattern can be replaced by a localized
density peak, sometimes exhibiting oscillating features, as depicted in Fig.~\ref{fig:drawing}(b). Sink is a heteroclinic solution
connecting two plane waves emitted by the sources at each side. 
The two waves collide at the sink position, where the incoming density currents are dissipated~\cite{Hohenberg_FrontsPulses}.

The sink density pattern is a result of the ability of the dissipative 
medium to smoothen out density ``dips'' and ``peaks'', which are
present in the standard interference pattern. When one of the incoming waves reaches the area occupied by the other, the resulting interference 
leads to decay of waves. Let us consider one of the simplest dissipative nonlinear wave equations, 
the complex Ginzburg-Landau equation~\cite{Hohenberg_FrontsPulses,Aranson_CGLEWorld} (CGLE)
\begin{align}
\frac{\partial \psi}{\partial t} &= \left[iD\frac{\partial^2}{\partial x^2} + iC|\psi|^2 - iB\right]\psi.
\label{eq:CGLE}
\end{align}
This equation can be obtained from the system of equations~(\ref{eq:PolaritonGPEWithA}) in the limit of fast 
relaxation time of the reservoir, which is ``slaved'' by the slower
$\psi$ dynamics, 
and in the linearized approximation $|\psi|^2\approx n_0\equiv {\rm Im} B/{\rm Im}C=(P-P_{th})/\gamma_C$ 
($n_0$ is the dynamical equilibrium density when the loss and gain are balanced). 
Here, we assumed a homogeneous pumping $P(x)=const>P_{\rm th}$ and introduced the threshold power $P_{\rm th} = \gamma_R \gamma_C / R$.
The CGLE parameters are $B=i\gamma_C/2 - (g_R+iR/2)(1+R n_0 / \gamma_A)P/\gamma_A$ and  $C=(g_R+iR/2)PR/\gamma_A^2-g_C$
with $\gamma_A=R n_0 + \gamma_R$.
It is clear that the existence  and stability of the homogeneous steady state 
with $n_0>0$ (which in general can be also a plane wave solution) 
requires Im$\,B>0$ and Im$\,C>0$. Under these conditions, perturbations of the steady state with density $n_0$
exponentially decay, which is the reason for the above mentioned smoothening. 
Any areas of density higher than $n_0$ correspond to net loss, and those with density lower than $n_0$ 
to net gain in Eq.~(\ref{eq:CGLE}).
Nevertheless, nontrivial (non-plane wave) stationary 
solutions can still exist~\cite{Hohenberg_FrontsPulses,Aranson_CGLEWorld,Xue_DarkSolitons}. 

One can formulate another necessary condition for stability of sinks based on the modulational (or Benjamin-Feir) stability
of the incoming plane waves~\cite{Hohenberg_FrontsPulses}. 
In the case of CGLE with real $D>0$, this is assured by the condition Re$\,C<0$, which in terms
of Eq.~(\ref{eq:PolaritonGPEWithA}) translates into $P/P_{\rm th}>(\gamma_C g_R/ \gamma_R g_C)$, as shown recently 
in~\cite{Ostrovskaya_DarkSolitons}. We note that this is a  necessary condition for stability, and an actual domain of stability
of plane waves may be smaller~\cite{Bobrovska_Stability}.

It may seem natural to treat sinks as dissipative analogues of bright solitons.
These states are, however, qualitatively different from each other.
Bright solitons exist in the conservative limit of the CGLE~(\ref{eq:CGLE}) with Im$\,(C,B,D)=0$, which is the
celebrated Nonlinear Schr\"odinger equation~\cite{Sulem} (or Gross-Pitaevskii equation in the context 
of degenerate bosons~\cite{PethickSmith}). 
Sinks require non-zero incoming currents for their existence~\cite{Hohenberg_FrontsPulses}, and exist in the case
of a stable background, Re$\,(CD)<0$, while bright solitons exist only in the self-focusing (modulationally unstable) case with $CD>0$.

\subsection{Sinks in the exciton-polariton model}

\begin{figure}
	\begin{center}
	\includegraphics[width=1.\columnwidth]{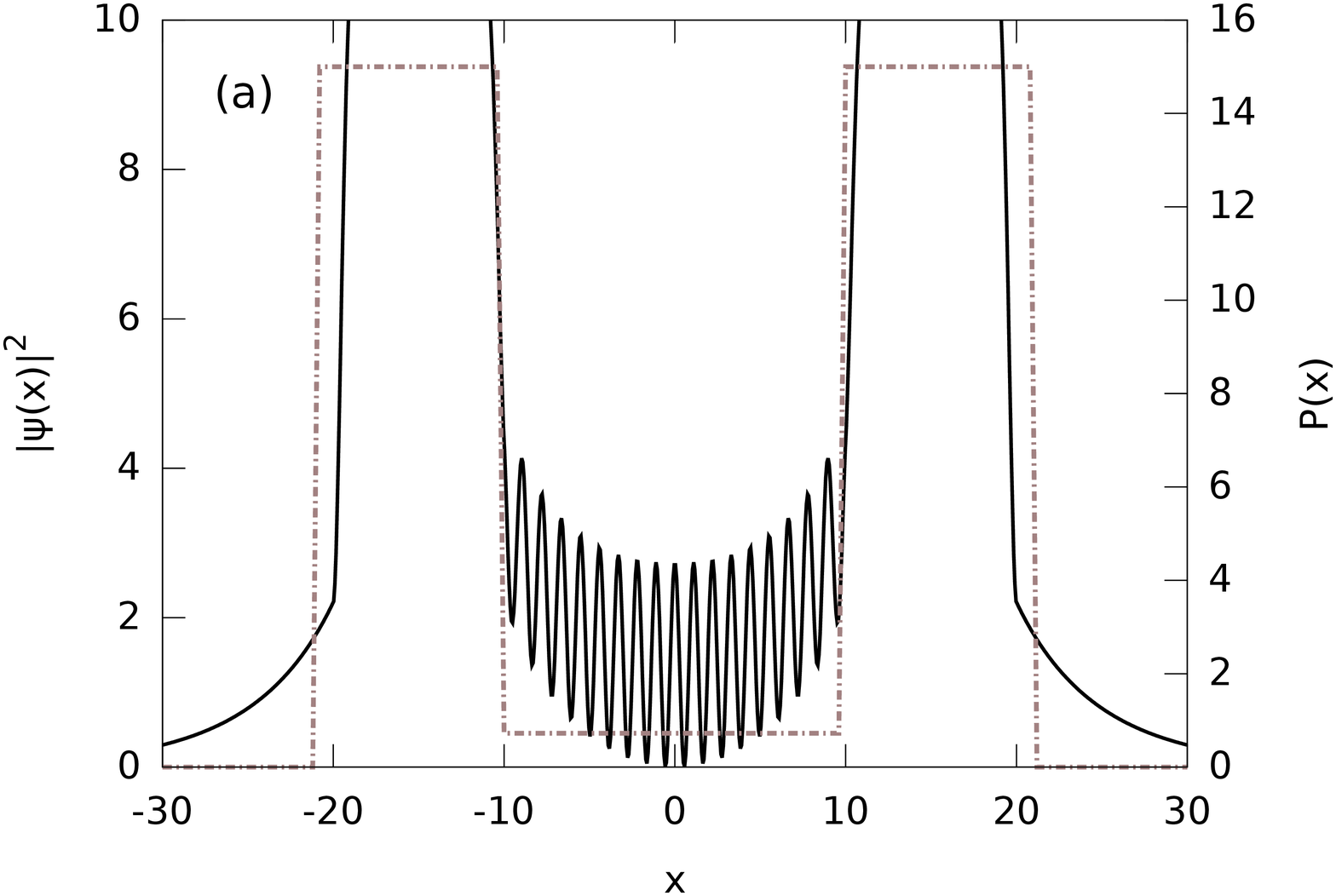}
		\includegraphics[width=1.\columnwidth]{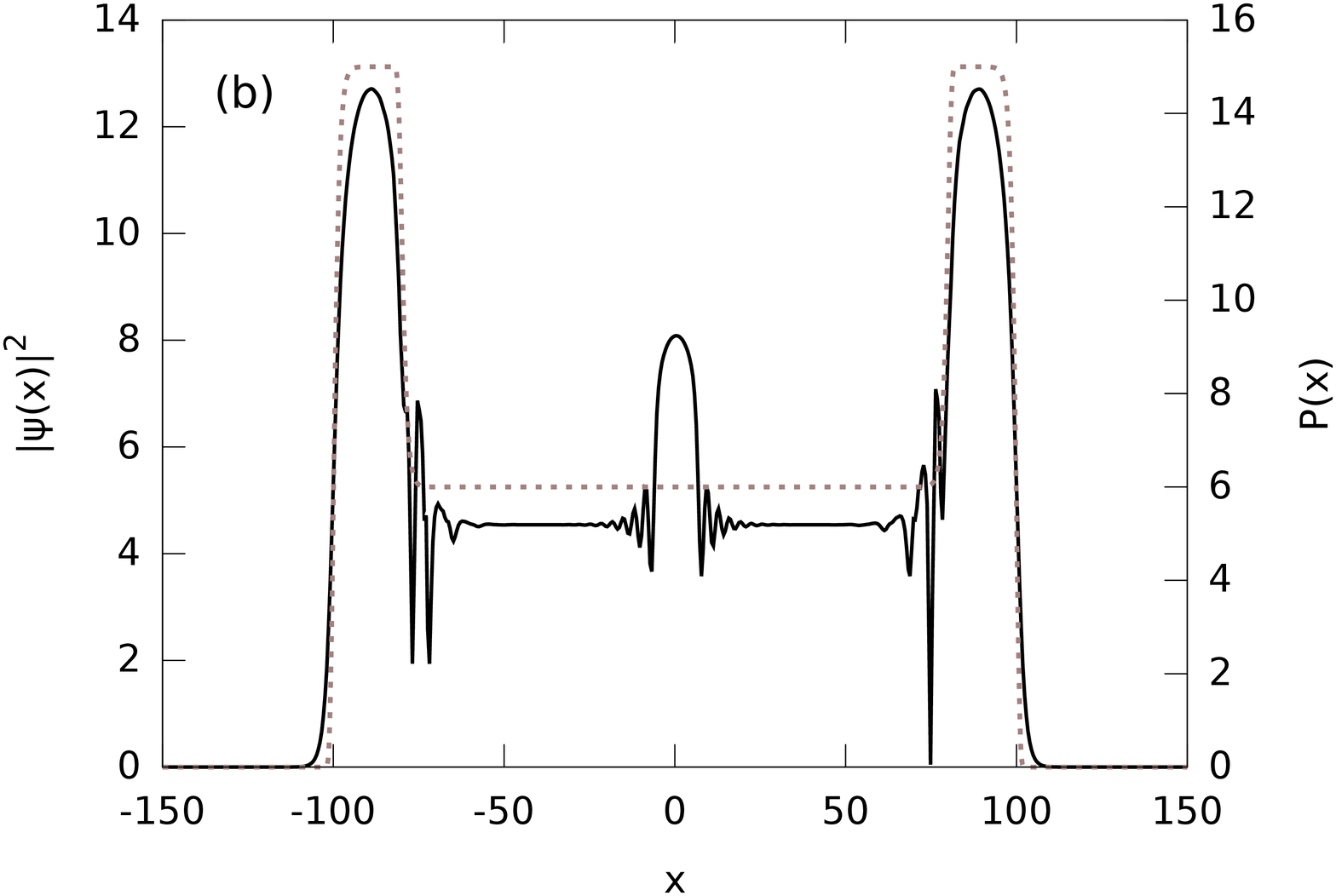}
	\caption{Patterns created by polariton waves emitted by high-intensity sources on the two sides 
with pumping profiles as in \eqref{eq:PProfile} (dashed lines).
Panel (a) presents a stationary interference pattern obtained by integration of~\eqref{eq:PolaritonGPEWithA} in time
for parameters $A=0.1$, $R=0.96$, $g_C=0.63$, $g_R=1.91$, $\gamma_C=0.9$, $\gamma_R=0.6$ at low pumping powers. 
(b) With a stronger pumping $P_0$
a nonlinear sink-type solution is created. 
Corresponding parameters in physical units are: time unit $\tau=\gamma_C^{-1}=3\,$ps, length unit $\xi=1.9\,\mu$m, $g=3.9\,\mu$eV$\mu$m$^2$, 
$R=9\times 10^{-3}\,\mu$m$^2$ ps$^{-1}$ for  $d=2\,\mu$m, $m^*=5\times 10^{-5} m_{\rm e}$, and $\beta=0.003$. 
} 
	\label{fig:InterfacePlusSink}
	\end{center}
\end{figure}

To create sink solutions in the model described by~(\ref{eq:PolaritonGPEWithA}), one has to provide sources of 
counterpropagating waves as described above. We consider the following pumping profile created by a pumping beam
with spatially varying intensity
\begin{equation}
\label{eq:PProfile}
P(x) = P_{\rm max}e^{-(x/w_b)^{\alpha}}-(P_{\rm max}-P_0)e^{-(x/w_s)^{\beta}}
\end{equation}
as shown in Fig.~\ref{fig:InterfacePlusSink} (dashed lines), where we used the smoothness parameters $\alpha=100$ and $\beta=80$.
The profile exhibits jumps of $P(x)$ at $x=\pm w_b$ and $x=\pm w_s$, typical of the Super-Gaussian terms in~(\ref{eq:PProfile}).

The sink is created in the central area with pumping intensity $P=P_0$. 
The side areas with $P=P_{\rm max}$ are the 
sources of polariton waves. This flow is obtained thanks to the repulsive polariton-polariton and reservoir-polariton
interactions $g_C$, $g_R>0$, which create an effective potential hill in the areas with high pumping density $P=P_{\rm max}$.
In result of interaction of the two nonlinear waves propagating towards the center, depending on the parameters of the system,
interference or localized sink patterns can appear, as shown in Fig.~\ref{fig:InterfacePlusSink}(a) 
and~\ref{fig:InterfacePlusSink}(b), respectively. Note that at the interface between the 
high- and low-pumping areas at $x=\pm w_s$ in~\ref{fig:InterfacePlusSink}(b),
oscillating time-dependent states form, which do not, however, preclude the formation of sinks.
We obtained also different, more complicated nonlinear patterns 
for other values of parameters, especially in the case when the two sources were relatively close to each other. 
These patterns did not have a localized character such as the one shown in Fig.~\ref{fig:InterfacePlusSink}(b).
A small value of relaxation coefficient $A=0.1$ was in some cases necessary to attenuate high-momentum modes in simulations 
and obtain physically relevant solutions.

\begin{figure}
	\begin{center}
	\includegraphics[width=1.\columnwidth]{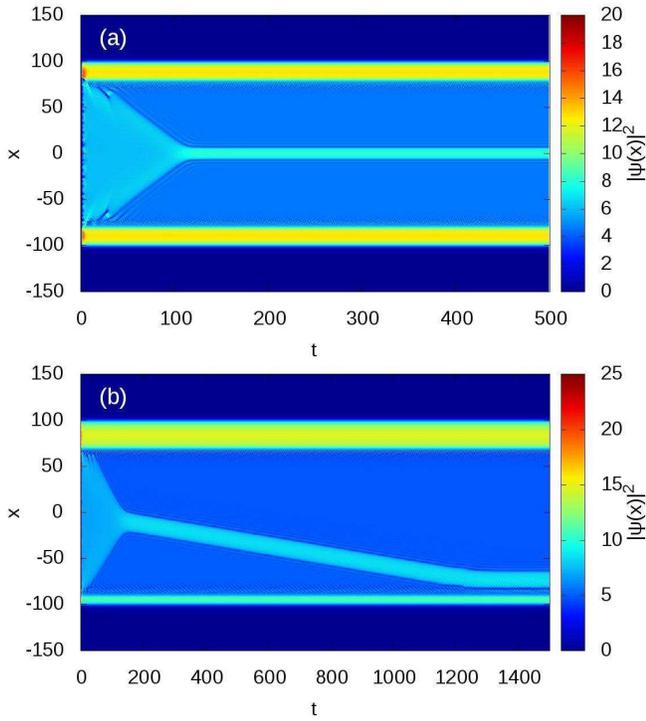}
	\caption{(a) Evolution towards a stationary sink state from a small initial noise. Parameters are as in Fig.~\ref{fig:InterfacePlusSink}(b). (b) The case of asymmetric pumping profile, with momentum mismatch between the waves from the two sources. 
The sink moves towards the weaker source, where it is stopped but not destroyed.}
	\label{fig:SinkCreation}
	\end{center}
\end{figure}

Figure~\ref{fig:SinkCreation}(a) shows the dynamics of the sink creation process. 
Initially, we assumed zero density of excitons and a small white noise in the polariton field, but we checked that the final stationary state
is practically independent of the form of the initial condition.
First, in the area between the sources a condensate is created with approximately zero momentum. 
The wave fronts generated by the sources gradually move towards the center, where they collide
creating the stable sink structure. The waves are ``stopped'' by the sink due to the nonlinear character of the gain
and dissipation. 

It is important to note that the sinks are in general not completely stationary, as in Fig.~\ref{fig:SinkCreation}(a),
but may be put in motion by the imbalance of momenta of the waves emitted by the two sources, see Fig.~\ref{fig:SinkCreation}(b). 
In this case the sink
moves with a constant velocity proportional to the mismatch between the two wavevectors~\cite{Hohenberg_FrontsPulses}. 
If the balance is restored after some period of time, the sink stops at the new position. The sink may be also stopped 
after reaching the weaker source, as demonstrated in Fig.~\ref{fig:SinkCreation}(b).

\section{Domain of existence and properties of sinks}

\begin{figure}
	\begin{center}
	\includegraphics[width=1.\columnwidth]{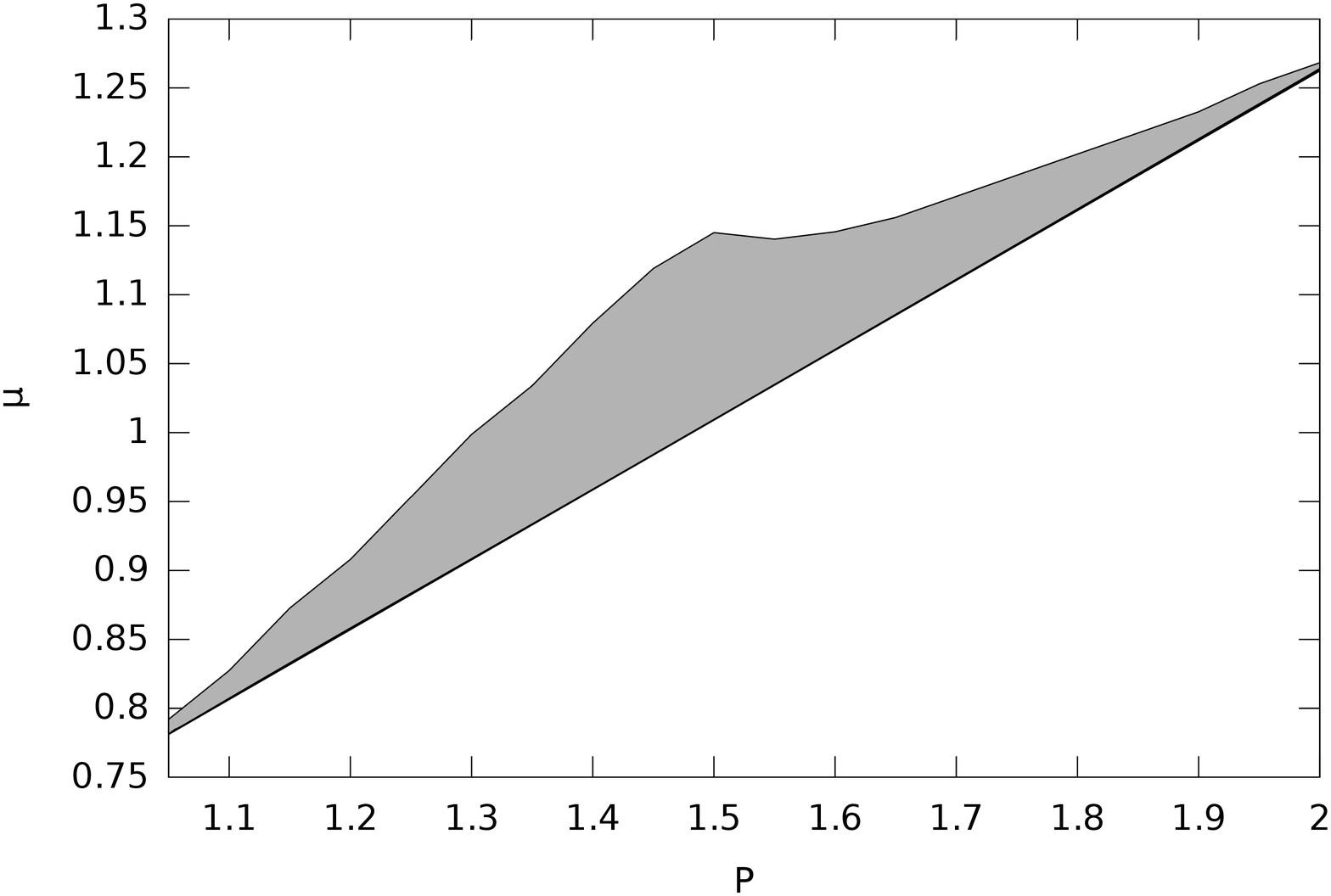}
	\caption{Phase diagram depicting parameters for which a sink solution was obtained with~\eqref{eq:StationaryGPE} (gray area). Parameters are $R=0.76,g_C=0.38,g_R=0.76,\gamma_C=0.75,\gamma_R=1$.}
	\label{fig:SinkMap}
	\end{center}
	
	\begin{center}
	\includegraphics[width=1.\columnwidth]{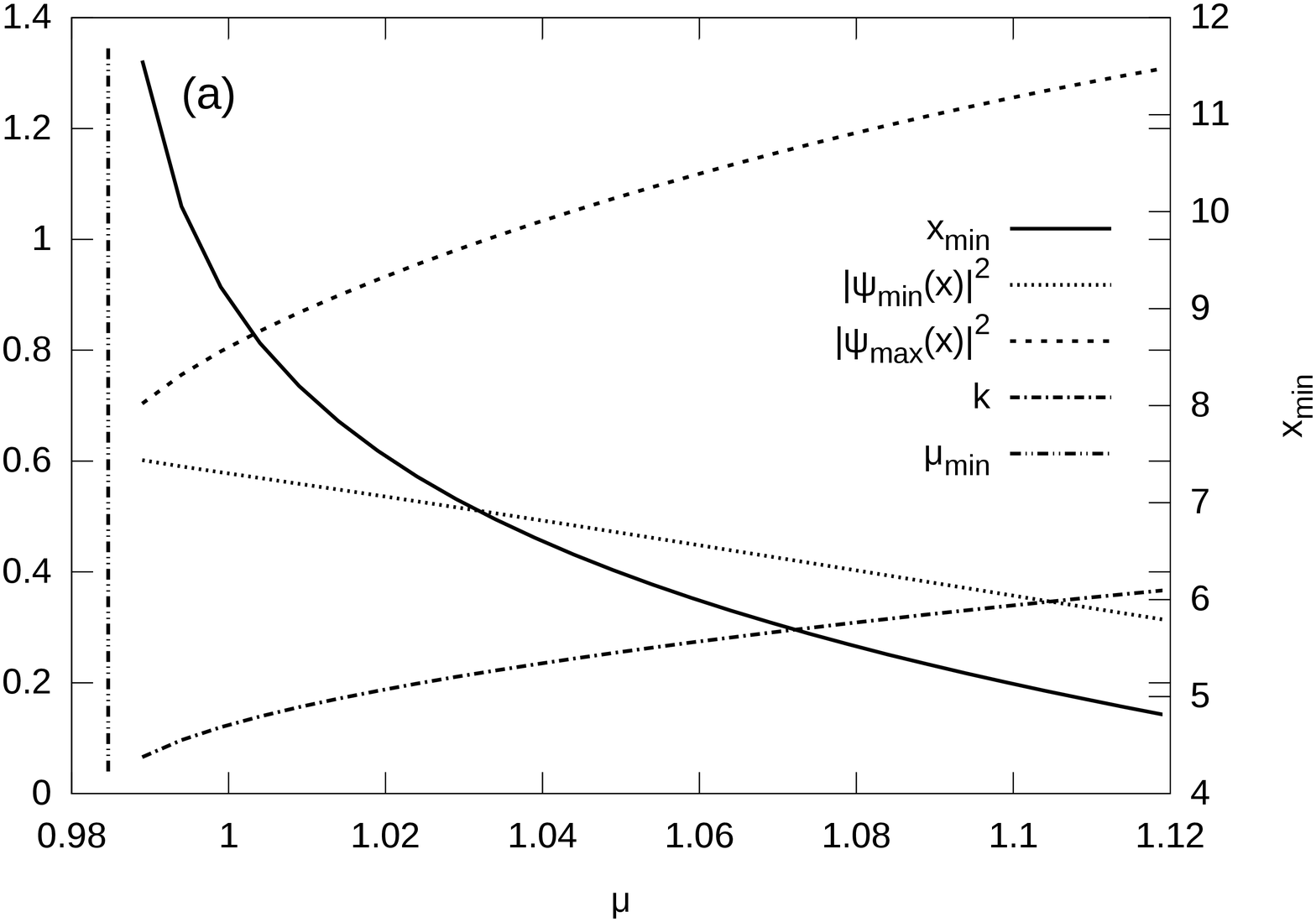}
	\includegraphics[width=1.\columnwidth]{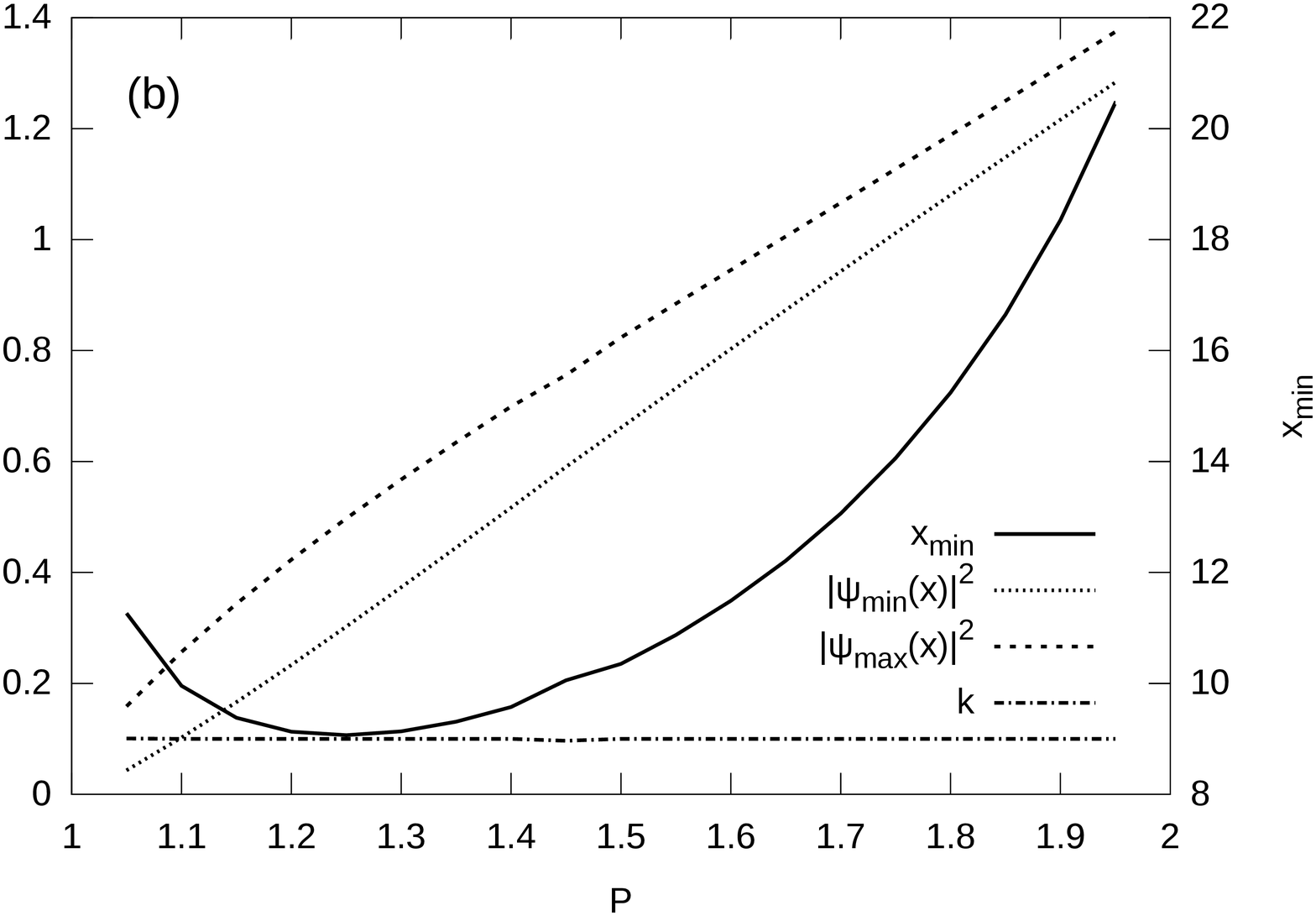}
	\caption{Panels show the values of $|\psi_{\rm min}(x)|^2$, $|\psi_{\rm max}(x)|^2$, $x_1$ and the wave vector $k$ 
(see Fig.~\ref{fig:drawing} for descriptions). On panel (a) we keep a constant value of $P=1.45$ but vary the chemical potential $\mu$. 
The value of $x_{\rm min}$ quickly decreases indicating that at higher $\mu$ the sink is more densely undulating. The dependence of $|\psi_{\rm min}(x)|^2$ and $|\psi_{\rm max}(x)|^2$ shows that the sink ``height'' is increased at higher $\mu$. In panel (b) we change the value of $P$ while $\mu$ 
is chosen slightly higher than the lower threshold $\mu_{\rm min}$ for which a sink solution occurs. 
The dependence of $|\psi_{\rm min}(x)|^2$, $|\psi_{\rm max}(x)|^2$ on $P$ indicates that sink height is approximately constant while the
background density $n_0$ grows with $P$.} 
	\label{fig:PConstMuMin}
	\end{center}	
\end{figure}

In this section we describe a systematic investigation of stationary sink solutions 
of the exciton-polariton model with homogeneous pumping, $P(x)={\rm const}$. 
We substitute $\psi(x,t) = \phi(x) e^{-i\mu t}$ and
$n_R(x,t)=n_R(x)$ into Eqs.~(\ref{eq:PolaritonGPEWithA}) to obtain a single ordinary differential equation
for the profile of the stationary state
\begin{align}
\frac{\textrm{d}^2 \phi}{\textrm{d} x^2}=&-\mu\phi+g_C|\phi|^2\phi+g_R\frac{P}{\gamma_R+R|\phi|^2}\phi + \nonumber\\
&+\frac{i}{2}\frac{RP}{\gamma_R+R|\phi|^2}\phi-\frac{i}{2}\gamma_C\phi
\label{eq:StationaryGPE}
\end{align}

We complement the above equation with boundary conditions. At $x=0$, which we chose to be the symmetry point without loss
of generality, the first derivative is equal to zero, $\textrm{d}\phi/\textrm{d}x=0$. At $x=+\infty$
the solution tends to the plane wave with the norm equal to $|\phi|^2=n_0=(P-P_{th})/\gamma_C$ 
(in practice we impose this condition on the last point of the computational mesh).

We solve the boundary problem with the shooting method using the Newton minimization algorithm. 
We keep $\frac{\textrm{d}\phi}{\textrm{d}x}|_{x=0}=0$ and change the value of $\phi(0)$ while solving~\eqref{eq:StationaryGPE} with the Runge-Kutta algorithm on a certain interval $0<x<x_{\rm max}$. The Newton method is then used to find a solution that satisfies boundary condition $|\phi|^2=n_0$ at $x=x_{\rm max}$ within a given tolerance. We then extend the interval boundary $x_{\rm max}$ slightly and repeat our procedure. This method proved to be an effective way to 
obtain a localized state on a large $x$ domain.

Figure~\ref{fig:SinkMap} presents a phase diagram showing the domain of existence of sink-type solutions
in the parameter space of $P$ and $\mu$ with values of other parameters fixed. The shaded area, corresponding
to parameters for which the algorithm converged to a sink solution, is limited from below by the
natural minimum given by the chemical potential of the steady state
\begin{equation}
\mu_{\rm min} = g_C n_0^2+\frac{Pg_R}{(\gamma_R + R n_0^2)}.
\label{eq:mumin}
\end{equation}
The range of $\mu$ for which the sink solution exists turned out to be the largest for moderate pumping intensities
$P\approx 1.5 P_{\rm th}$. Although the range of $\mu$ appears to be small, it corresponds in fact to a broad
range of wavenumbers of incoming waves, from approximately zero to around 6, 
as shown in Fig.~\ref{fig:PConstMuMin}(a) with a dash-dotted line.
In this figure, the vertical line corresponds to the minimal value of $\mu$, given by Eq.~(\ref{eq:mumin}).
The dependence between the chemical potential and the wavenumber of incoming waves can be calculated by taking
the $x\rightarrow\pm \infty$ limit away from the sink, for which Eq.~(\ref{eq:StationaryGPE}) reduces to 
$\mu=k^2+g_C n_0^2+Pg_R/(\gamma_R + R n_0^2)$ under the condition of balanced gain and loss.
It is then clear that the necessary condition for the existence of sinks is that the kinetic energy is much smaller
than the nonlinear energy of the wave. 

The dashed and dotted lines in Figure~\ref{fig:PConstMuMin}(a) depict the minimum and maximum density of the sink profile,
see Fig.~\ref{fig:drawing}. With increasing $\mu$, which corresponds to increasing $k$ vectors of incoming waves,
the sinks become larger and more highly modulated, while in the $\mu=\mu_{\rm min}$ limit they transform smoothly to 
the flat homogeneous state $\phi(x,t) = n_0 e^{-i\mu_{\rm min}t}$. The position of the first minimum of the density $x_{\rm min}$ (solid line)
decreases with the increase of $\mu$, which is related to the increasingly dense interference pattern of the tails of 
the sink.

Figure~\ref{fig:PConstMuMin}(b) shows the dependence of the same properties of the sink as described above, but with
increasing pumping power $P$. Here $\mu$ is kept at a slightly increased level with respect to $\mu_{\rm min}$. This corresponds
to a constant small value of $k$. In this case, the increase of $P$ leads to the increase of average sink density,
but without almost any change of the difference $|\phi_{\rm max}|^2-|\phi_{\rm min}|^2$. On the other hand, the value
of $x_{\rm min}$ shows a strongly nonmonotonous character, decreasing for small $P$ and increasing again for large $P$.
This shows that in the case of small $k$ the position of first minimum is not simply related to the incoming wavevector $k$.

\section{Analytical solution}

\begin{figure}
	\begin{center}
	\includegraphics[width=0.7\columnwidth]{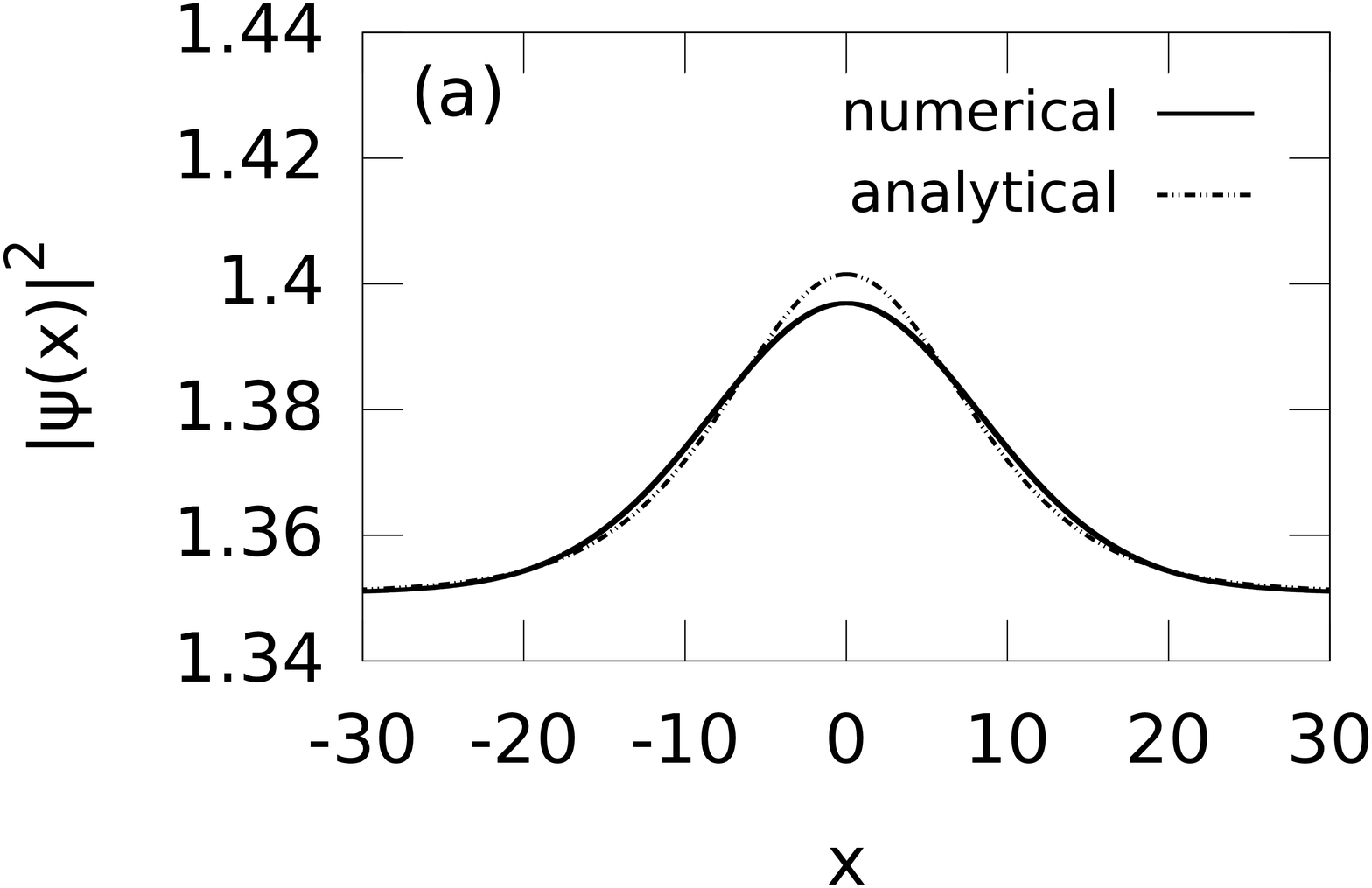}
	\includegraphics[width=0.7\columnwidth]{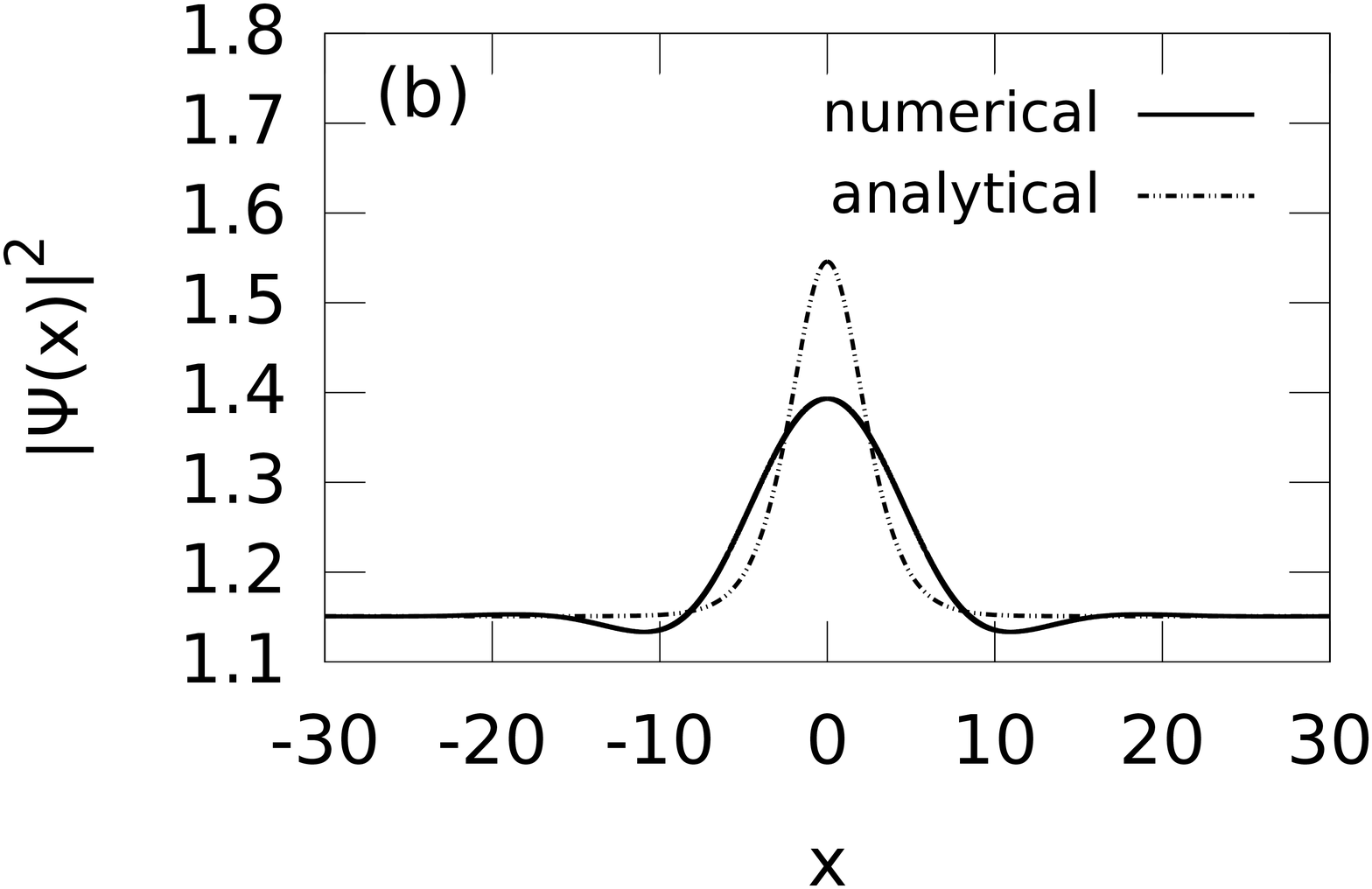}
	\caption{Comparison of the numerical sink solution from the shooting method with the approximate analytical solution for two sets of parameters. Parameters are $R=0.76$, $g=0.38$, $g_R=2g$, $\gamma_C=0.75$, $\gamma_R=1$, $P=2.0$, $\mu=1.2683$  for (a) and $P=1.85$, $\mu=1.2173$ for (b).}
	\label{fig:analytical}
	\end{center}
\end{figure}

If the ``height'' of the sink is relatively small compared to the steady state density, $||\psi(x,t)|-n_0| \ll n_0$,
and the amplitude $|\psi(x,t)|$ is slowly varying in space, 
an approximate analytical solution can be found~\cite{Popp_FromDarkSolitons}.
We rewrite the steady-state solution in a homogeneously pumped condensate applying the Madelung transformation for $\psi(x,t)$
\begin{equation}
\begin{split}
&\psi(x,t)=a(x) e^{i(\varphi(x)-\mu t)},\\
&n_R(x,t) = n_R(x),
\label{steady-state}
\end{split}
\end{equation}
where $a(x)$ is the amplitude, $\varphi(x)$ is the phase and $\mu$ is the chemical potential of the condensate. 
Neglecting the spatial derivatives of $a(x)$, Eq.~(\ref{eq:PolaritonGPEWithA}) can be rewritten as 
\begin{eqnarray}
\label{phi}
\nonumber
i\mu&=&i\left[ga^2(x)+g_Rn_R(x)+\left(\frac{{\rm d}}{{\rm d}x}\varphi(x)\right)^2\right]+\\
&-&\left[\frac{1}{2}\left(Rn_R(x)-\gamma_C\right)
-\frac{{\rm d^2}}{{\rm d}x^2}\varphi(x)\right]
\end{eqnarray}
The real part of Eq.(\ref{phi}), which can be interpreted as the continuity equation, gives
\begin{equation}
\label{n1}
n_R(x)=\frac{2}{R}\left(\frac{\textrm{d}^2}{\textrm{d}x^2}\varphi(x)\right)+\frac{\gamma_C}{R}
\end{equation}
On the other hand, solving the equation for reservoir density~(\ref{eq:PolaritonGPEWithA}) in the steady state we get
\begin{equation}
n_R(x)=\frac{P}{a^2(x)R+\gamma_R}.
\end{equation}
We expand this formula into Taylor series of degree two around $a^2(x)=n_0=(P/\gamma_C)-(\gamma_R/R)$
\begin{equation}
\label{n2}
n_R(x)=-\frac{\gamma_C\left(a^2(x)R\gamma_C-2PR+\gamma_C\gamma_R\right)}{PR^2}
\end{equation}
Comparing Eqs.~(\ref{n1}) and (\ref{n2}) we obtain
\begin{equation}
\label{AA}
a^2(x)=-\frac{2PR\left(\frac{\textrm{d}^2}{\textrm{d}x^2}\varphi(x)\right)-PR\gamma_C+\gamma_C\gamma_R}{R\gamma_C^2}
\end{equation}
From the imaginary part of Eq.~(\ref{phi}), using~(\ref{n1}) and~(\ref{AA}) we get
\begin{eqnarray}
\label{mu_large}
\nonumber
\mu&=&-\frac{g}{R\gamma_C^2}\left[2PR\left(\frac{\textrm{d}^2}{\textrm{d}x^2}\varphi(x)\right)-PR\gamma_C+\gamma_C^2\gamma_R\right]+\\
&+&g_R\left[\frac{2}{R}\left(\frac{\textrm{d}^2}{\textrm{d}x^2}\varphi(x)\right)+\frac{\gamma_C}{R}\right]+\left(\frac{\textrm{d}}{\textrm{d}x}\varphi(x)\right)^2.
\end{eqnarray}
With the definition $\xi(x)=\frac{\rm{d}}{\rm{d}x}\varphi(x)$ we obtain the first order differential equation
\begin{eqnarray}
&&2\left(\frac{g_R}{R}-\frac{gP}{\gamma_C^2}\right)\frac{\rm d}{{\rm d}x}\xi(x)+\xi^2(x)+\\
&&+\left[g\left(\frac{P}{\gamma_C}-\frac{\gamma_R}{R}\right)+\frac{g_R\gamma_C}{R}-\mu\right]=0
\end{eqnarray}
With the solution
\begin{equation}
\label{xi}
\xi(x)=\frac{\sqrt{\delta}}{R\gamma_C}\tan\left(\frac{1}{2}\frac{\sqrt{\delta}}{\alpha}\right),
\end{equation}
where $\delta=R(PRg\gamma_C-R\mu\gamma_C-g\gamma_C^2\gamma_R+g_R\gamma_C^3)$ and $\alpha=gPR-g_R\gamma_C^2$.
Calculating the amplitude with~(\ref{AA}) we finally obtain
\begin{eqnarray}
a^2(x)&=&-\frac{1}{R\gamma_C^2}\left[\frac{\delta P}{\alpha}\left\lbrace1+\tan^2\left(\frac{1}{2}\frac{\gamma_C\sqrt{\delta}}{\alpha}x\right)\right\rbrace+\right. \nonumber\\
&-&\left. PR\gamma_C+\gamma_C^2\gamma_R\right]
\label{eq:analytical}
\end{eqnarray}

Comparison between the analytical and numerical solution obtained with the shooting method is shown in 
Fig.~\ref{fig:analytical}. In general, very good agreement is obtained for small $k$, when the sink profile
is flat and broad, with no  oscillating tails. The tails obviously cannot be reproduced by the approximate 
solution~(\ref{eq:analytical}), which is visible in the (b) panel of Fig.~\ref{fig:analytical}.

\section{Two-dimensional case}

\begin{figure}
	\begin{center}
	\includegraphics[width=1.\columnwidth]{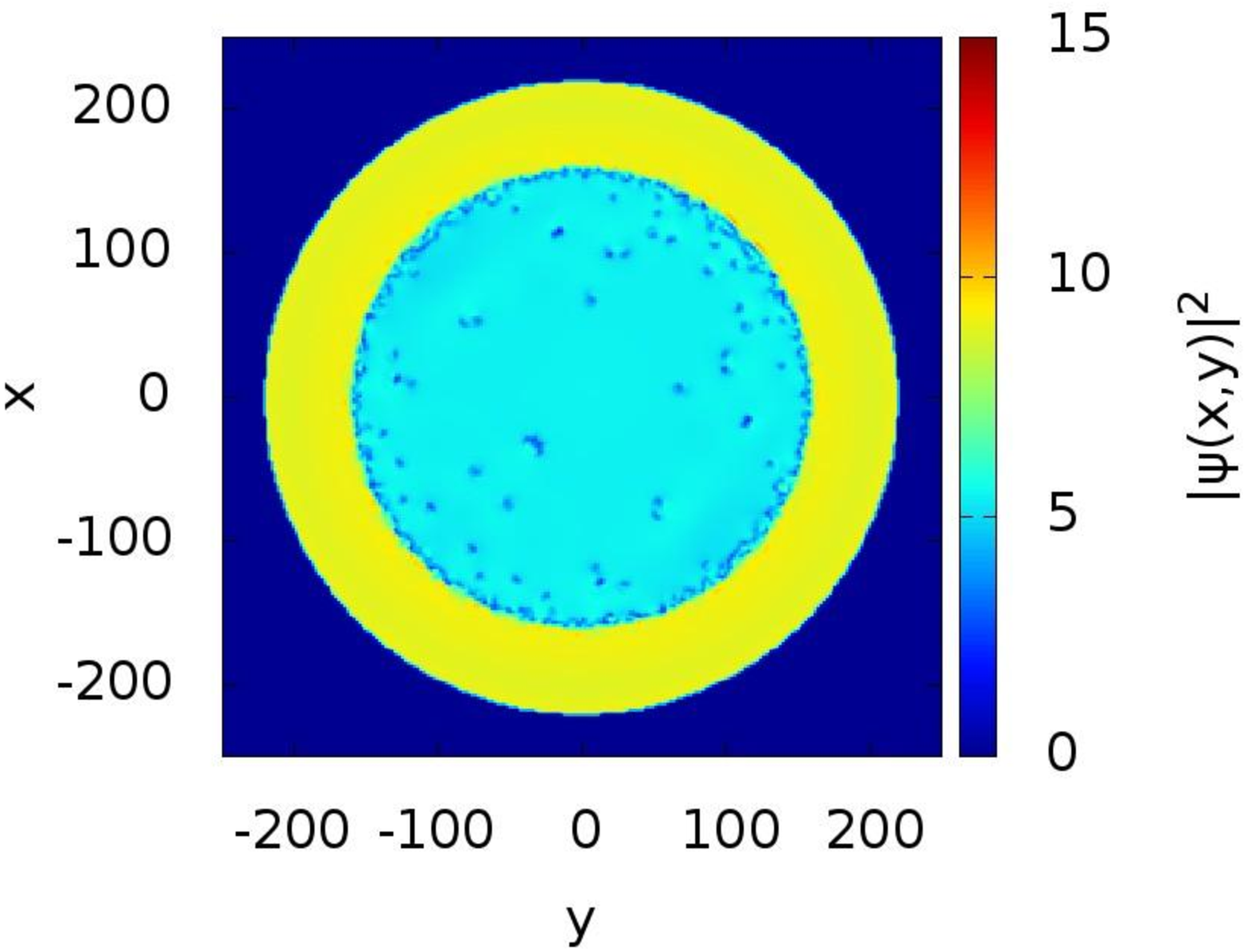}
	\includegraphics[width=1.\columnwidth]{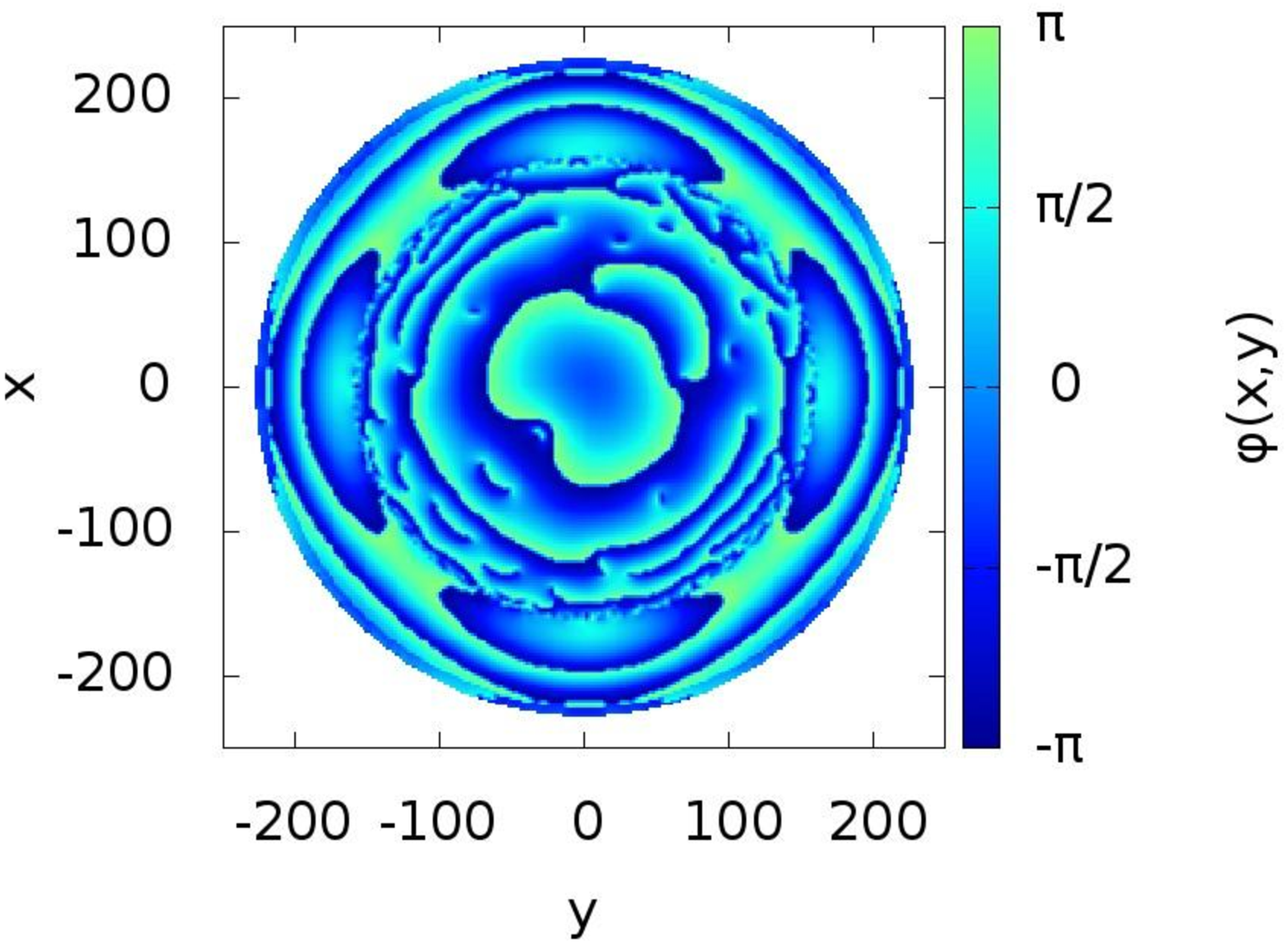}
	\caption{Snapshot of a two-dimensional solution of \eqref{eq:2DPolaritonGPEWithA} after a long time of evolution $t=1600$, generated by a pumping profile in the shape of a ring. Quantum vortices are clearly visible in the density (top) and phase (bottom) 
of $\psi(x,y,t)$. Sink solutions in 2D are absent due to proliferation of vortices. Parameters are $A=0.1$, $R=0.96$, $g_C=0.63$, $g_R=1.91$, $\gamma_C=1$, $\gamma_R=0.6$ with $P_0=6$, $P_{\rm max}=10$,$r_1=155$, $r_2=220$.}
	\label{fig:2dPhase}
	\end{center}
\end{figure}

Additionally, we performed a series of simulations to investigate whether sink creation is possible in the two-dimensional (2D) version
of the exciton-polariton model, described by the equations
\begin{align}
i\frac{\partial \psi}{\partial t} &= \bigg[-D\left(\frac{\partial^2}{\partial x^2} + \frac{\partial^2}{\partial y^2}\right) +g_C|\psi|^2 +g_Rn_R + \nonumber\\
&+\frac{i}{2}\left(Rn_R-\gamma_C\right)\bigg]\psi, \\
\frac{\partial n_R}{\partial t} &= P(x,y)-\left(\gamma_R+R|\psi|^2\right)n_R,
\label{eq:2DPolaritonGPEWithA}
\end{align}
where $g_{R,C}$, $\gamma_{R,C}$, $R$, and $P(x,y)$ are dimensionless parameters obtained from physical ones in an analogous way as in the
1D case. 

One of possible choices of the pumping profile corresponds to a constant $P=P_0$ pumping intensity on a 
circle of radius $r_1$, surrounded by a ring of $P=P_{\rm max}$ with inner and outer radius $r_1$ and $r_2$, respectively.
In Figure~\ref{fig:2dPhase} we show a typical state obtained with this kind of pumping profile and an evolution from a small initial noise.
In general, after a certain time of evolution, smaller or lager number of vortices are spontaneously created, and often a stationary
state could never be reached even with a very long integration time. 
Vortices may appear spontaneously during condensation in a process analogous to the Kibble-Zurek 
mechanism~\cite{KibbleZurek,Deveaud_VortexDynamics,Matuszewski_UniversalityPolaritons}
as well as due to the emergence of supercurrents in an inhomogeneous system~\cite{Berloff_VortexLattices,Deveaud_QuantumVortices}.
Despite using various combinations of system parameters, as well as asymmetric ring pumping profiles,
we were not able to obtain any stable structures that would resemble
one-dimensional sinks of the previous sections. 

We note that similar pumping profiles were used in several experiments~\cite{Bramati_VortexChains,Ring_Patterns} where
multi-lobe or vortex patterns were observed. However, the experimental patterns were in most cases regular, which suggests that they correspond to the linear regime as in Fig.~\ref{fig:InterfacePlusSink}(a). In the case of strong nonlinear interactions, regular vortex chain 
patterns could be observed with the resonant pumping scheme \cite{Bramati_VortexChains}. 
Depending on the system parameters, these could be destroyed by spontaneously nucleating vortices created through a hydrodynamic instability, 
which is consistent with our simulations. 

\section{Conclusions}

We demonstrated the existence and stability of a family of bright sink solutions of the open-dissipative
polariton model. In contrast to bright solitons of conservative models, sinks exist in the case of repulsive 
interactions and are created in a collision of counter-propagating waves.
We studied the dynamics of sink formation in a realistic one-dimensional polariton model with appropriately chosen pumping profile. 
We studied the domain of existence of sinks in parameter space and their physical properties. 
An approximate analytical formula for the sink shape, valid in the case where sinks do not possess oscillating tails, was determined.
In the two-dimensional ring-shaped confguration, sink solutions were not found due to the spontaneous appearance of vortices.

\acknowledgements

We acknowledge support from the National Science Center grant DEC-2011/01/D/ST3/00482.

\end{document}